\documentclass[11pt]{article}

\usepackage{times}
\usepackage{a4}
\usepackage{verbatim}
\usepackage{graphicx}
\usepackage{amssymb}
\usepackage{amsmath}
\usepackage{xspace}
\usepackage[affil-it]{authblk}

\usepackage{natbib}
\bibpunct{(}{)}{;}{a}{,}{,}
\newcommand{\df}{\ensuremath{\textrm{d}}\xspace}
\newcommand{\dd}{\ensuremath{\ \textrm{d}}\xspace}
\newcommand{\aL}{\ensuremath{\boldsymbol{\alpha}_L}\xspace}
\newcommand{\bL}{\ensuremath{\boldsymbol{\beta}_L}\xspace}
\newcommand{\aU}{\ensuremath{\boldsymbol{\alpha}^U_L}\xspace}
\newcommand{\bU}{\ensuremath{\boldsymbol{\beta}^U_L}\xspace}
\newcommand{\Lab}{\ensuremath{L_{\alpha \beta}^U}\xspace}

\title{Testing the Multiverse: Bayes, Fine-Tuning and Typicality}

\author{Luke A. Barnes%
  \thanks{Electronic address: \texttt{L.Barnes@physics.usyd.edu.au} \\
	  Proceedings of the Philosophy of Cosmology UK/US Conference, 12th - 16th September 2014, Tenerife, Spain}}
\affil{Sydney Institute for Astronomy \\
School of Physics, University of Sydney \\
NSW 2006, Australia}

\begin{document}
\maketitle

\section{Introduction}
Theory testing in the physical sciences has been revolutionized in recent decades by Bayesian approaches to probability theory. Here, I will consider Bayesian approaches to theory extensions, that is, theories like inflation which aim to provide a deeper explanation for some aspect of our models (in this case, the standard model of cosmology) that seem unnatural or fine-tuned. In particular, I will consider how cosmologists can test the multiverse using observations of this universe.

Cosmologists will only ever get one horizon-full of data. Our telescopes will see so far, and no further. At any particular time, particle accelerators reach to a finite energy scale and no higher. And yet, it would be an unnatural constraint on our theories for them to fall silent beyond the edge of the observable universe and above a certain energy. Natural, simple theories need not confine themselves to the observable. How do we speculate beyond current data?

In particular, how do we evaluate (what I will call) theory extensions? That is, physical theories whose main attraction is that they provide a deeper, more natural understanding of some effective theory. For example, the appeal of cosmic inflation is its natural explanation of some of the ``initial conditions'' of the standard model of cosmology. The postulates of the standard model --- a homogeneous and isotropic Robertson-Walker (RW) spacetime, a set of energy components and their densities (matter, radiation and a cosmological constant), and an initial set of adiabatic, Gaussian density and tensor perturbations --- can explain all (or almost all) the cosmological data at our disposal: the expansion of the universe, big bang nucleosynthesis, the angular power spectrum of the cosmic microwave background (CMB), the galaxy and Lyman alpha forest power spectra, the baryon acoustic oscillation (BAO) scale, the luminosity distance-redshift relation of Type 1a supernovae, and more.

So, why not simply declare cosmology to be finished? We have a model that explains all the data. Consider the following kind of reason for extending our cosmological theory. In the standard model of cosmology, photons in the CMB that are separated in the sky by more than $\sim$1 degree were scattered by patches gas that have never been in causal contact with each other. And yet the entire CMB is at the same temperature, to one part in 100,000. If, alternatively, we propose that there was a period of accelerating expansion in the very early universe, then the regions we see in the CMB have been in causal contact, allowing them to come to thermal equilibrium. And thus, inflation solves the \emph{horizon} problem, so the standard story goes.

Note well: the horizon problem does not involve a theory failing to predict an observation. Theories never predict their initial conditions. Rather, we argue that something about our model is open to a deeper explanation because it is unnatural, improbable, or an unexplained coincidence.

Examples could be multiplied. General Relativity explains what to Newtonian gravity was a bare postulate: the equivalence of inertial and gravitational mass. Supersymmetry doesn't currently explain any data, but would explain why quantum corrections do not drive the Higgs Boson mass to the Planck scale, a fact which would otherwise be highly unnatural.

A calculation is required to make these arguments robust. Returning to inflation: how probable is an isotropic CMB given inflation, and not given inflation? And how simple is inflation as a hypothesis, given that we don't know what the inflaton is? How generic (probable?) are the initial conditions that lead to inflation? Observations can tell us something about the initial conditions of the observable universe; when should we accept a dynamical theory of those initial conditions, rather than simply postulating them?

Can we attack these questions with probability theory at all? Cosmology promises to stretch our interpretation of probabilities. It will be my contention here that objective Bayesian probabilities provide a consistent framework for extrapolating cosmological theories beyond our universe, and isolate the pertinent questions to ask of such theories.

\section{Objective Bayesian Probability}

\subsection{Probability from Uncertainty}

We will start with an (oversimplified) overview of probability, and in particular my impressions of how it is used in the physical sciences. The interpretation of probability has a long and surprisingly turbulent history. In one corner stands the frequentists, for whom probabilities measure the relative frequencies of events in hypothetical infinitely repeated trials (or, for \emph{finite} frequentists, in actual, known trials). When a scientist wants to test their ideas, they calculate the probability of the data given the theory. If this probability (known as a \emph{likelihood}) passes certain tests, then we can announce that the theory is not disconfirmed.

The mathematical foundation of this approach was provided by \citet{Kolmogorov1933}, who builds probability theory from mathematical axioms, independent of any particular application to statistics. Probability, like tensor calculus or conic sections, is a tool that may or may not be useful to the scientist in the investigation of some physical system.

If probabilities are frequencies of outcomes, it makes no sense to ask for the probability \emph{of} a theory. We cannot compare the number of universes that obey Newtonian gravity with the number that obey Einstein's General Relativity. This is not a criticism of frequentism by its opponents. Ronald Fisher, the patron saint of frequentism, stated that ``we can know nothing of the probability of hypotheses or hypothetical quantities'' \citep{Fisher1921}.

In the other corner stands the Bayesians\footnote{It is a simplification to speak of just two corners, but sufficient for our purposes.}. The basis of this approach is not abstract axioms but an attempt to start from the \emph{desiderata} of rationality and develop probability theory as generalized logic. While classical logic is concerned with what follows deductively --- if $A$ then $B$ --- probability theory will include weaker degrees of certainty --- if $A$ then probably $B$. Probabilities such as $p(B | A)$ (``the probability of $B$ given $A$'') quantify the degree of certainty of the proposition $B$ given the truth of the proposition $A$. Classical logic's implication $A \rightarrow B$ is the special case $p(B|A) = 1$; those two are the same statement. The goal is not merely to quantify subjective degrees of belief, that is, the psychological state of someone who believes $A$ and is considering $B$. Just as classical logic's $A \rightarrow B$ says nothing about whether $A$ is known by anyone, but instead denotes a connection between the truth values of the propositions $A$ and $B$, so $p(B|A)$ quantifies a relationship between these propositions\footnote{Neither are we considering degrees of truth; $A$ and $B$ are in fact either true or false.}. 

How should degrees of certainty be assigned to certain propositions? \citet{Jaynes2003} invites us to imagine a reasoning robot: insert a \emph{given} proposition $A$ in one slot, and the proposition of interest $B$ in the other slot, and out comes a number indicating the degree of certainty.  We program the robot according to the following desiderata:
\begin{itemize}  \setlength{\itemsep}{-2pt}
\item[D1.] Probabilities are represented by real numbers. This ensures that degrees of plausibility can be compared on a single scale.
\item[D2.] Probabilities change in common sense ways. For example, if learning C makes B more likely, but doesn't change how likely A is, then learning C should make AB more likely.
\item[D3.] If a conclusion can be reasoned out in more than one way, then every possible way must lead to the same result.
\item[D4.] Information must not be arbitrarily ignored. All given evidence must be taken into account.
\item[D5.] Identical states of knowledge (except perhaps for the labeling of the propositions) should result in identical assigned probabilities.
\end{itemize}
Perhaps surprisingly, these desiderata are enough. Cox's theorem (\citet{Jaynes2003} and \citep{Caticha2009} are required reading) shows that quantities assigned according to these desiderata obey the same rules as probabilities. In particular, we have a rule for each of the Boolean operations `and' ($AB$), `or' ($A+B$) and 'not' ( $\bar{A}$),
\begin{align}
p(AB|C) &\equiv p(A|BC) ~p(B|C) \equiv p(B|AC) ~p(A|C) \label{eq:prod}\\
p(A+B|C) &\equiv p(A|C) + p(B|C) - p(AB|C) \label{eq:sum} \\
p(\bar{A}|C) &\equiv 1 - p(A|C) ~. \label{eq:not}
\end{align}
These are \emph{identities}, holding for any propositions $A$, $B$ and $C$ for which the relevant quantities are defined. In particular, from Equation \eqref{eq:prod} we can derive Bayes' theorem,
\begin{equation} \label{eq:bayes}
p(A|BC) = \frac{p(B|AC) ~ p(A | C)}{p(B | C)} ~.
\end{equation}
Bayes's theorem often comes attached to a narrative about `prior' probabilities, which depend only on `known' `background' information (or worse, \emph{temporally} prior information), that is updated with new `data' to produced revised `posterior' probabilities. None of this is essential to Bayesianism.

\emph{The} goal of Bayesian probability theory is to calculate the probability of the proposition of interest $A$, given everything we know $K$. If you are handed $p(A|K)$ from the clouds, then your work is done. If, however, $p(A|K)$ is too much to handle then you'll have to break it into smaller pieces. In particular, the sum total of everything you know $K$ is likely to be expressible as a conjunction, $K = BC$, in which case Bayes's theorem is very useful. We use probability identities to write probabilities we \emph{want} in terms of probabilities we \emph{know}.

\subsection{The Rise of Bayesianism}

A revolution in the physical sciences over the last few decades has transformed what we do with data. New methods have been advanced because of a fundamental change in the way that scientists view probability. From these new foundations have come a new approach and a new set of tools, all marching under the banner of Bayes.

To underscore the dominance of Bayesian probability theory, a recent NASA Astrophysics Data System (ADS) search of the astronomy and physics literature for articles with the word ``Bayesian'' or ``Bayes'' in the title returned 7555 papers. A search for ``frequentist'' or ``frequentism'' in the title returned 71 papers, half of which also have ``Bayes'' in the title. Most of these are comparing methods. Frequentist methods are still used, and will not always be advertised as such. Nevertheless, this does show how few physicists and astronomers advertise their methods as frequentist. I have never seen frequentism defended in a scientific paper. On the rare occasions that the word appears, it is usually as a synonym for ``oversimplified'' or ``archaic'' or ``wrong''.

Why has Bayesianism risen so quickly in the physical sciences? I think that there are two main reasons.

\begin{figure}[t]
\centering
    \begin{minipage}{0.5\textwidth}
       \includegraphics[width=\textwidth]{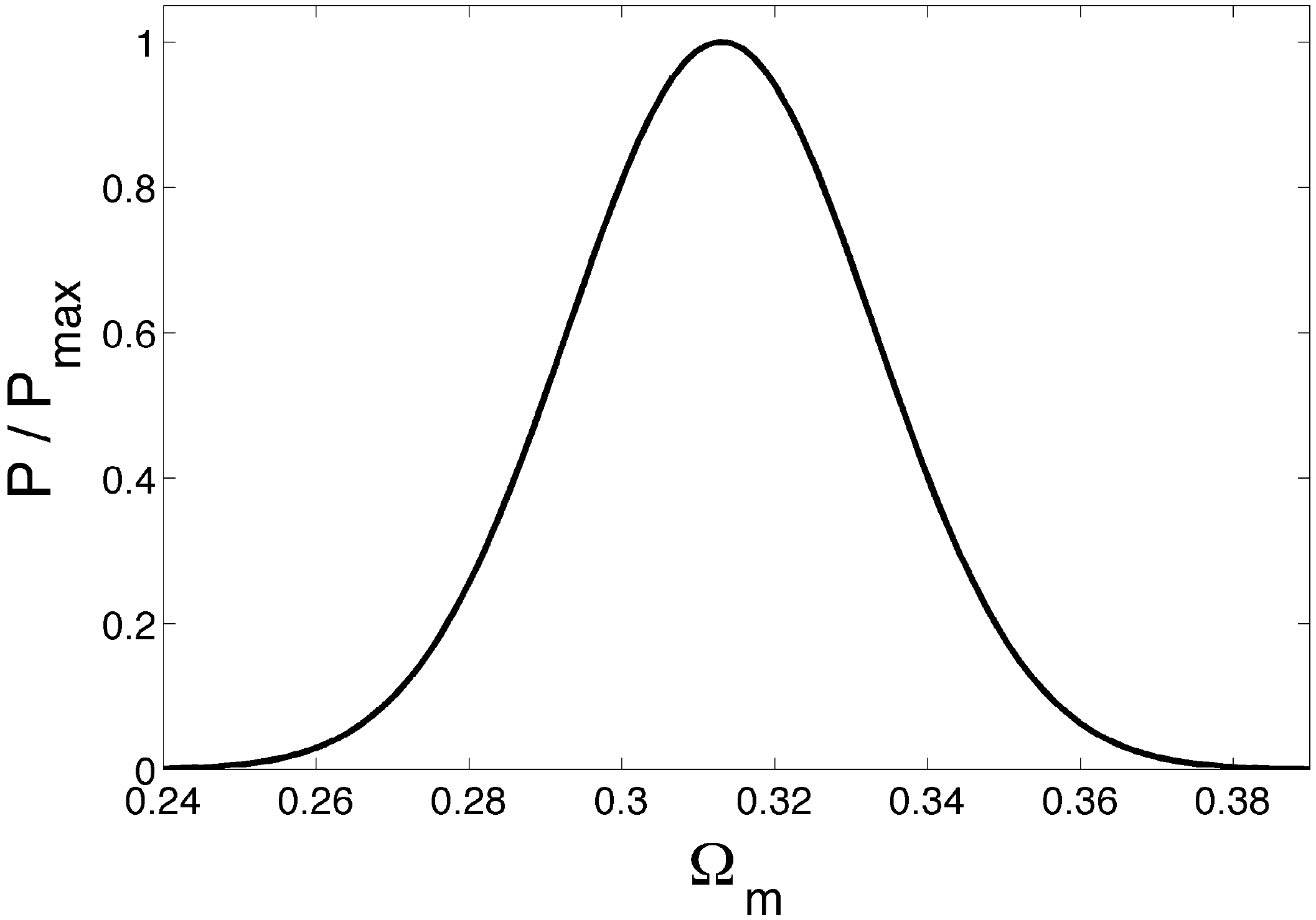}
    \end{minipage}
    \hspace{0.05\textwidth}
	\begin{minipage}{0.4\textwidth}
	    \caption{Constraints from the Planck CMB satellite on the average cosmic density of matter. The $y$-axis shows the probability (density) of a particular value of the parameter, normalized to the maximum value.} \label{fig:Omega_m}
	\end{minipage}
\end{figure}

Firstly, Bayesianism makes good sense of theory \emph{testing}. Figure \ref{fig:Omega_m} shows the constraints from data from the Planck CMB satellite \citep{Planck2015} on the average cosmic density of matter, relative to the critical density. The $y$-axis shows the probability (density) of a particular value of the parameter, normalized to the maximum value.

What \emph{exactly} does the $y$-axis quantify? It isn't a finite or hypothetical frequency --- it's not saying that $\sim$95\% of universes we polled (or would hypothetically poll) have a mass density parameter between 0.27 and 0.36. The width of the peak is not an indication of the range of matter densities in different regions of the universe. It is not a chance, as if the density of the universe is a stochastic property that every third Sunday of the month is less than 0.27. The universe only has one value of its average density, and so knows of only one point on the $x$-axis.

The $y$-axis of this plot most plausibly quantifies our degree of certainty. And yet, this is not a subjective credence. The Planck data analysis team is not reporting the effect that their satellite's instruments have had on their state of mind. What this plot reports is the implications of cosmological data for the knowledge of cosmological parameters.

More generally, science must be able to conclude, for example, that quantum mechanics is more likely to correctly describe atoms than classical electromagnetism. (Otherwise, what's the point? We'd never learn anything.) This probability must be a statement about propositions, about states of knowledge. It cannot be a statement about frequencies or chances, because it isn't a statement about the universe at all, or even a hypothetical ensemble of universes. Nature knows nothing of our incorrect theories.

This does not mean that frequencies and chances are useless. A frequencies is a useful way to describe data. Chances are legitimate postulates of a physical theory, for example in describing the macroscopic state of a thermodynamic system or the indeterminacy of quantum systems. Bayesian probability theory does not imply that quantum probabilities are epistemic, or that statistic mechanics needs only human ignorance to link microphysics with thermodynamics. Rather, the claim is that frequencies and chances are insufficient for testing theories.

Secondly, the practice of Bayesian statistics exhibits a deep clarity and unity. The methods of orthodox statistics are a grab-bag of techniques, each intuitively reasonable but without any deeper insight into which is the best, or even of what ``best'' should mean. For example, \citet{Jaynes2003} reports that, faced with linear regression (with both variables subject to an error of unknown variance), the orthodox textbook of \citet{Kempthorne1971} formulates sixteen different methods, and, being unable to choose between them, concludes with ``It is all very difficult." A later survey of orthodox methods can ``give only a long, somewhat dreary, list of one \emph{adhockery} after another, with no firm final conclusions.'' In contrast, the Bayesian approach gives the scientist the impression of asking the right questions of the data, with no hidden assumptions and no black boxes.

\subsection{Has Bayesianism Succeeded?}

The claim of the Bayesian is that there are objective degrees of certainty or \emph{credences} that can be modelled as probabilities. They are neither frequencies (actual or hypothetical), chances, nor merely subjective.

The reader might, and probably (!) should, be skeptical as to whether such an ambitious quantification of reasoning has indeed been achieved by the Bayesians. It might seem like alchemy, turning the base metal of ignorance into the gold of a precise probability distribution. Keep in mind, however, that Bayesian probabilities do not imply statistical frequencies: it does not follow from $p(B|A) = 0.5$ that there is a population of $A$'s that we could sample half of whose members are $B$'s.

Further, Bayesian probabilities do not quantify everything that $A$ says about $B$. Suppose that a mystery black box will flip a coin. What is the probability of heads $H$, given this information ($A$)? The Bayesian has no reason to prefer one side to the other; in particular, the coin and/or box might be biased towards one side, but we don't know which. To reflect this ignorance, we assign $p(H|A) = 0.5$. Now suppose that we examine the coin and box, and discover that the coin is (as best we can tell) perfectly symmetric and unbiased, and inside the box we find a mechanism that has shown no evidence of bias in the last billion flips. What is the probability of heads $H$, given this new, detailed information $B$? It hasn't changed: $p(H|B) = 0.5$. Should the Bayesian be worried that the probability does not reflect the vast difference in the information in $A$ and $B$? Should we seek to expand probability to take into account this difference, using fuzzy probabilities or assigning distributions rather than numbers? Perhaps. But the unchanged probability is in some sense the right answer. Sure, we've learned a lot about the coin and the box, but this knowledge shouldn't have changed our belief that heads will turn up\footnote{It will, appropriately, change the probability that the coin is biased, given a sequence of flips. Given $A$, a series of repeated heads will quickly convince us that the coin is biased. Given $B$, we will resist such a conclusion for longer, believing in the light of our examination of the coin and box that the repeated heads are mere chance.}.

The assignment of probabilities is not derailed by ignorance. Ignorance is a state of knowledge, and probabilities describe states of knowledge. It may seem like assigning $p(H|A) = 0.5$ using the principle of indifference is misleadingly precise. We should reserve definite probability assignments for cases like $B$, and should instead say of $A$ that ``I don't know". But this would sell ourselves short. ``I don't know which one of these \emph{two} statements is true" is a very different state of knowledge from ``I don't know which one of these \emph{trillion} statements is true". Our probabilities can and  should reflect the size of the set of possibilities; the principle of indifference is invoked as a special case when this size is all we have. The assigned probabilities are only misleadingly precise if overinterpreted.

Nevertheless, Bayesian probability theory is not without worries. Some are pseudo-problems, such as the ``problem of old evidence'' \citep{Glymour1980}\footnote{Exercise for the reader. Hint: $p(E|B) = 1$ does not follow from ``I know $E$".}. More troubling is the assignment of prior probabilities. Recall that prior probabilities are simply probabilities calculated using less than everything we know. So the problem is really: how do we assign probabilities when we don't know very much? The problem of the prior is particularly acute when faced with a continuum of possibilities, such as a probability distribution over a variable. We cannot say that each value is equally probable, or that each interval in an infinite range is equally probable, since these distributions do not sum (integrate) to one. The probabilities are worryingly shuffled by a change in variable. How do we model ignorance of an infinite number of possibilities?

Various methods have been advanced to solve this problem, including Jeffrey's prior and Jaynes et al's Principle of Maximum Entropy. Whether these are successful is beyond the scope of this paper, but their failure would not sink Bayesianism. It would leave an open problem in the program. The most that the Bayesian might have to give up in light of these worries is that probabilities can be assigned to any proposition given any state of knowledge. For example, it seems absurd to suppose that there is such a thing as the probability that "the toilet paper is purple" given that "the plate is orange"\footnote{Thanks to Eric Winsberg for this example.}, that there is some \emph{number} that uniquely captures the relationship between those propositions.

Faced with infinite possibilities, or vague statements about purple toilet paper, we might have to refrain from assigning a probability until more information is given. \citet{Jaynes2003} argues that the problem of infinities is similar to the problem of vague statements --- we haven't really specified the problem until we know the limiting procedure that generates the infinity. Where one should draw the "too vague" line, however, is not clear.

\section{Extending the Laws of Nature (as we know them)}

\subsection{Taking Stock}
We want apply Bayesian probability theory to the extension of the laws of nature, and then in particular to the multiverse. First, we must take stock of the laws of nature as we know them. We consider the somewhat idealized case in which we have identified the effective laws of nature that govern the physical regimes relevant to our observational evidence. Let,
\begin{itemize}  \setlength{\itemsep}{-2pt}
\item $U$ = our observations of this universe.
\item $B$ = everything else we know. 
\item $L$ = the laws of nature as we know them.
\end{itemize}

$U$ represents the sum total of our observations of this universe, including every telescope observation and every experiment. $B$ represents everything else we know, such that $UB$ represents everything we know. $B$ includes mathematical knowledge, and in particular all of theoretical physics. A statement such as ``a bound test particle moving according to Newton's law of gravity would obey Kepler's laws of planetary motion'' is true even if no particles actually obey Newton's law. It is not a statement about the actual world.

Regarding $L$, I'm thinking here of the Lagrangian of the standard model of particle physics plus general relativity, but the details won't much matter. In a typically entertaining footnote, David Griffiths imagines ``that God has a giant computer-controlled factory, which takes Lagrangians as input and delivers the universe they represent as output'' \citep[p. 373]{Griffiths2008}.

Actually, we need more than just the functional form of the Lagrangian. The equations of the laws of nature --- as we know them --- contain free parameters, numbers which are not predicted by the theory itself, but without which the laws are not fully specified. In addition, it is the \emph{solutions} to the equations that describe a possible universe. We require further parameters to specify a particular universe from amongst the family of possible universes. These are usually specified as \emph{initial conditions}, or more generally, boundary conditions.

We will represent the free parameters of the laws of nature, referred to as the \emph{constants of nature}, as the set of numbers \aL. Similarly, we will represent the initial conditions required to specify a solution/universe by\footnote{The notation can be easily extended to functions or more advanced mathematical structures than lists of numbers.} \bL. The subscript $L$ is a reminder that it is only in the context of a particular theory that a measurement of our universe becomes a \emph{fundamental} constant.

We wish to evaluate the probability of our theory $L$, given the evidence we have $p(L | UB)$. We use Bayes' Theorem:
\begin{equation}
p(L | UB) = \frac{p(L|B)}{p(U|B)} ~ p(U | LB) ~ .
\end{equation}
However, $L$ is missing its parameters, and will not predict quantities until they are specified. We can introduce the free parameters $\aL,\bL$ as \emph{nuisance} parameters, to be integrated out:
\begin{equation} \label{eq:postab}
p(L | UB) = \frac{p(L|B)}{p(U|B)} ~ \int p(U | \aL \bL L B) p( \aL \bL | L B) \dd\ \aL \dd \bL ~.
\end{equation}
A few points to note. The first term on the top is the `prior' probability of the law $L$, $p(L|B)$. This is the probability that $L$ describes this universe, given no information about this universe. Here is the place to formalize and implement Occam's razor --- we expect simpler theories to be more probable \citep[the interested reader is encouraged to consult][Chapter 28]{MacKay2003}.

The first term inside the integral (the likelihood) is where the theory, equipped with the appropriate constants and initial conditions, shows its predictive powe by predicting observations. The second term inside the integral is the prior probability of the free parameters, that is, the probability of the parameters falling into a certain range, given no information about this universe. Note that this term takes $L$ as given --- the parameters have no law-independent meaning.

Our observations of the universe not only constrain $L$ but its free parameters. We can, with a slight abuse of notation\footnote{Specifically, there are two abuses of notation. We are using $\aU$ and $\bU$ to refer to parameter regions, whereas before $\aL$ and $\bL$ referred to particular values. Secondly, we should be placing propositions into our probability functions. We can think of $\aU$ as representing the proposition ``the value of the fundamental constants of the theory $L$ lie in the region $\aU$''.}, denote by $\aU$ and $\bU$ the \emph{set} of free parameters consistent with experiment, such that,
\begin{equation}
p(\aU \bU | LUB) \gg p(\overline{\aU \bU} | LUB) ~.
\end{equation}

Our goal as physicists is to identify the laws of nature that govern our observations of the universe. Ideally, $L$ describes our observations better than any rival theory, $p(L | UB) \gg\ p(\bar{L} | UB)$, and while there exist a range of candidate deeper theories into which $L$ could be embedded, none is significantly preferred by our data. We do not assume that $L$ is \emph{the} ultimate law of nature.

\subsection{Why Extend the Laws?}

\begin{figure}[t]
\centering
       \includegraphics[width=\textwidth]{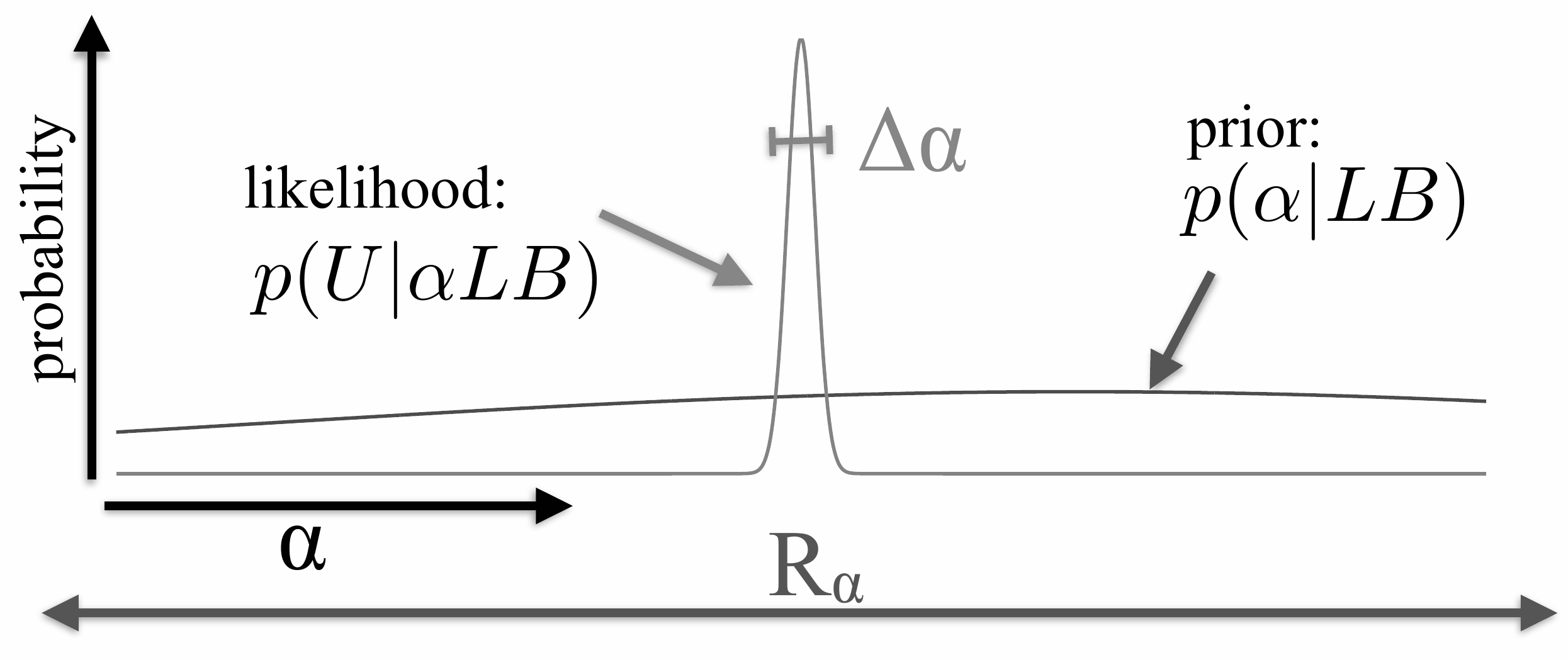}
	    \caption{A Bayesian picture of a fine-tuned theory. $p(U | LB) = \int p(U | \alpha L B) ~ p( \alpha | L B) ~ \textrm{d} \boldsymbol\alpha \sim \frac{\Delta \alpha}{R_\alpha} \ll 1$.} \label{fig:FTBayes}
\end{figure}

One particular way in which we would like a deeper physical theory to differ from current theories is with regard to the constants of nature. In particular, we want them gone, and we can see why from Equation \eqref{eq:postab}. A sharply-peaked $p(U | \aL \bL L B)$ as a function of the free parameters (\aL, \bL) is precisely what physicists usually mean by ``fine-tuned'' --- if the theory only adequately explains the data for a very narrow range of its free parameters, then we are suspicious. To illustrate in the one-dimensional case (Figure \ref{fig:FTBayes}), suppose that the prior $p(\alpha | L B)$ is non-zero over a range $\sim R_\alpha$, and the likelihood $p(U | \alpha L B)$ is sharply peaked in a range of values $\Delta \alpha$, and negligible outside. (Remembering that the prior is normalized over $\alpha$, but the likelihood isn't.) Then when we integrate over the nuisance parameter $\alpha$, $p(U | L B) \sim \Delta \alpha / R_\alpha$. Unless the prior probability is fortuitously peaked in the same range, the likelihood $p(U|LB)$ will be very small.

The discovery that a theory is fine-tuned opens the door for an alternative theory to replace it. This theory could have a broader likelihood, a narrower prior, or have no free parameters at all. Note that a preference for such a theory is not merely aesthetic, nor simply the desire to summarize the behaviour of nature as succinctly as possible.

\subsection{How to Evaluate an Theory Extension}
So, we seek an extension to the laws of nature in which fewer arbitrary constants appear. How does the Bayesian  evaluate theory extensions?

Consider, as an example, a detective entering a crime-scene. She relies on background evidence $B$ (what she knew before she entered the room), and inside the room she collects evidence $E$. The evidence clearly indicates that $K$, a local thug, is the killer: $p(K|EB) \gg p(\bar{K} | EB)$. Still, there may be puzzling or suspicious aspects of the hypothesis $K$; perhaps $K$ didn't know the victim. We thus are led to consider other propositions; not rival theories, but extensions to $K$. We might wonder whether $K$ was ($C$) contracted to kill the victim by a local mob boss. We can evaluate this extended hypothesis in light of the data as follows:
\begin{equation}
p(CK|EB) = p(C | KEB) p(K|EB) ~.
\end{equation}
Now, we suppose that $C$ doesn't explain the evidence of the crime-scene beyond the hypothesis $K$, $p(E | CKB) = p(E | KB)$. That is, $C$ seeks to explain $K$, and $K$ explains $E$. For example, $K$'s fingerprints at the scene are not rendered more or less probable by his status as a \emph{contract} killer. We can then write,
\begin{equation}
p(CK|EB) = \frac{p(K | CB)}{p(K | B)} p(C|B) p(K|EB) ~.
\end{equation}
There are three factors of interest here. The first fraction denotes the probability of $K$ being the killer given the contract hypothesis (and $B$), relative to the probability of $K$ being the killer given background information alone. This is where the theory-extension shows its worth, by leading us to expect that $K$ would kill the victim. The second term is the prior probability of $C$, $p(C|B)$; the theory $C$ is penalized if it is implausible given the background information. Thirdly, $p(K|EB)$ is the posterior probability of $K$, which by hypothesis is close to one.

\subsection{Extending the laws of nature}

Consider an extension to the laws of nature. We consider a deeper theory $T$, which aims to explain the laws and constants of nature as we observe them $L \aU \bU$. (For convenience, we will write $\Lab \equiv L \aU \bU$ to denote the whole ``laws + parameters'' package.). We assume that this deeper theory does not explain the data we observe beyond its ability to explain $L$, that is, $p(U | T \Lab B) = p(U | \Lab B)$. For example, let $\Lab$ be the standard model of cosmology, beginning just prior to nucleosynthesis, and let $T$ be inflation, which ends well before nucleosynthesis. Our prediction of the statistical properties of the CMB needs only $\Lab$; inflation does not predict the properties of the CMB \emph{beyond} predicting the ``initial conditions" of the standard model of cosmology.

The formalism is then analogous to the crime-scene case above:
\begin{equation}
p(T \Lab|UB) = \frac{p(\Lab | TB)}{p(\Lab | B)} p(T|B) p(\Lab|UB)~.
\end{equation}
We can expand the fraction above,
\begin{equation} \label{eq:pTLab}
p(T \Lab|UB) = \frac{p(\aU \bU | TLB)}{p(\aU \bU | LB)} \frac{p(L| TB)}{p(L | B)} p(T|B) p(\Lab|UB) ~.
\end{equation}
This is similar to the Bayesian formalism by which theories are tested with data, except that we are testing the theory extension  $T$ by using the effective theory and its measured constants $\Lab$ as if they were \emph{data}. Equation \eqref{eq:pTLab} highlights three questions to ask of \emph{any} proposed extension to the laws of nature as we know them. Firstly, given the theory $T$, the effective laws of nature $L$ and background information $B$, how probable are the constants and initial conditions of our universe? Secondly, given the theory $T$ and background information $B$, how probable are the effective laws of nature $L$? Finally, given background information $B$, how probable is the theory $T$?

Let's look at some ways to do away with free parameters.

\section{Extension 1: Replace Free Parameters with Mathematical Constants}

To some, free parameters are a call to action, a hot poker in the Bayesian posterior. We are not satisfied, and we will not be satisfied until every physical measurement can be predicted from theory alone.  \citet{Einstein1949} dreamed of a set of equations such that ``within these laws only rationally completely determined constants occur (not constants, therefore, whose numerical value could be changed without destroying the theory)''.

In our formalism, this theory would set $p(\Lab | TB)= 1$: given the deeper theory, there is only one low-energy effective theory with only one possible value of each ``constant". Measuring the constants of nature would be akin to drawing a circle and determining its radius and circumference in order to ``measure'' $\pi$.

Unifying scientific theories can reduce the number of free parameters in physics. For example, Maxwell's unification of electricity and magnetism  showed that $c = 1/\sqrt{\epsilon_0 \mu_0}$ ($c$ speed of light,  $\epsilon_0$ vacuum permittivity, $\mu_0$ vacuum permeability), thus reducing the number of free parameters of physics by one. This is a step in the right direction, but the progress of science can just as easily increase the number of constants by, for example, discovering a new fundamental particle.

Einstein's dream is not without its worries. A ``perfect", unity likelihood is often a clue that the theory is ad hoc or jerry-rigged. For example, a theory with a large number of siblings --- that is, mutually exclusive but similar theories that are equally probable given our background information --- will only receive a small slice of the total prior probability of the family. This is, in essence, why theories with free parameters are suspicious in the first place. The theory can be thought of as a large family of theories, one for each value of the free parameter. 

Thus, we need to worry about the prior probability of our deeper theory $p(T|B)$. It may have no free parameters, but if it is but one member of a large set of similar theories, the prior probability may still be small. In particular, while by hypothesis we cannot vary the parameters of the theory, this may merely indicate that we must look for fine-tuning at the next level deeper, as it were. Varying the effective parameters of our laws may require varying the deeper theory, leaving us no less at the mercy of a large set of possibilities.

This highlights one of Steven Weinberg's wishes in ``Dreams of a Final Theory'' \citep{Weinberg1993}, which he calls \emph{logical isolation}. Weinberg argues that, while quantum mechanics is not logically inevitable, ``any small change in quantum mechanics would lead to logical absurdities'' (p. 70). In this sense, there is no obvious continuum of theories, of which quantum mechanics is just one. The Bayesian argument above fits nicely with Weinberg's intuition. Total logical isolation, however, seems too much to ask. Mathematical consistency is not trivial, but neither is it a rarity. There is no ultimate equation of our physical universe to which we can hope to say ``mathematically, that's how things \emph{must} be''.

In addition, a theory that requires no initial conditions, or that somehow predicts its own initial conditions, would be rather strange. Rather than specifying the dynamical properties of physical objects in the form of counterfactuals, it would   specify the state of the universe. For example, a Newtonian version of such a theory would not state that if two masses ($m_1$, $m_2$) are separated by distance $r$, then they would experience a force with magnitude $Gm_1 m_2 / r^2$. Rather, it would specify position as a function of time $\boldsymbol{r}(t)$ for each particle in the universe. Rather than the complexity of the phenomena of the universe giving way to simple fundamental laws, a theory with no initial conditions would seem to require complexity all the way down.

\section{Extension 2: Replace Free Parameters with Dynamical Entities}

We have expounded the ingredients of physical theories as we know them: laws, constants and initial (or boundary) conditions. The laws describe dynamical entities --- fields, particles, spacetime etc. So, one way in which a constant could disappear in a deeper theory is by changing identity to become a dynamical quantity. The fine-structure constant, for example, could be the local value of a field. We can test this hypothesis by looking for changes in the value of the fine-structure constant over cosmic time and cosmic distances. To date, no convincing variation has been found \citep{2011PhRvL.107s1101W,2012MNRAS.422.3370K,2012arXiv1207.6223C,2015MNRAS.447..446W}.

Two problems immediately arise. Firstly, if the fine-structure constant is replaced by a quantum field, then it seems that we have merely replaced one constant with the parameters that describe the field. (In fact, a field that varies so slowly over the observable universe requires a very low mass.) Secondly, even if we could replace our constants with a totally constant-free field, this doesn't seem like progress. We have replaced a single number with a function: an infinite collection of numbers, one attached to each spacetime point. If we are in a typical place in the universe, then there is no further rationale for the value of the ``constant'' that we observe. There is some function that varies across spacetime, and we happen to be in the part that has $\alpha \approx 1/137$.

\subsection{The Fine-Tuning of the Universe for Intelligent Life}

However, there are good reasons to believe that we are not in a typical place in the universe. The universe is not an experiment. We are not Dr. Frankenstein, setting up our equipment, choosing the initial conditions, and observing the setup at our leisure. We are the monster --- we have awoken in a laboratory and are trying to figure out how it made us. Not all rooms can create a monster, so the fact that we are observing at all is a very stringent constraint on the contents of the room.

Similarly, not all laws of nature can be \emph{scientific} laws, because not all laws of nature create scientists. There are certain equations that will not be written on a chalkboard in any universe that they describe. If the evolution of conscious observers shows a strong preference for certain laws or certain regions of parameter space, then an explanation for the values of the constants naturally arises. The reason why this set of constants exists at all is that there are a sufficiently large number of universe domains, with enough variation in their properties that at least one of them would hit on the right combination for life. The reason why we observe that we are in one of these rare regions is that we couldn't be anywhere else.

Beginning in the 1970's, a number of physicists have noticed the extreme sensitivity of the life-permitting qualities of our universe to the values of many of the physical constants and cosmological parameters of our universe. Seemingly small changes in the free parameters of the laws of nature as we know them have dramatic, uncompensated and detrimental effects on the ability of the universe to support the complexity needed by physical life forms. I have elsewhere reviewed the scientific literature on the fine-tuning of the universe for intelligent life \citep{2012PASA...29..529B}. Here are a few examples.
\begin{itemize}
\item The existence of structure in our universe \emph{at all} places stringent bounds on the cosmological constant. Compared to the range of values for which our theories are well defined --- roughly $\pm$ the Planck scale --- the range of values that permit gravitationally bound structures is no more than one part in $10^{110}$.
\item A universe with structure also requires a fine-tuned value for the primordial density contrast $Q$. Too low, and no structure forms. Too high and galaxies are too dense to allow for long-lived planetary systems, as the time between disruption by a neighbouring star is too short. This places the constraint $10^{-6} \lesssim Q \lesssim 10^{-4}$ \citep{1998ApJ...499..526T}.
\item The existence of long-lived stars, which produce and distribute chemical elements and are a stable source of energy that can power chemical reactions, requires an unnaturally small value for the ``gravitational coupling constant'' $\alpha_G = m^2_\textrm{proton} / m^2_\textrm{Planck}$; or, equivalently, that the proton mass be orders of magnitude smaller than the Planck mass. For stars to be stable at all, we require $\alpha_G \lesssim 10^{-33}$ \citep{2008JCAP...08..010A}.
\item The existence of any atomic species and chemical processes whatsoever places tight constraints on the relative masses of the fundamental particles and the strengths of the fundamental forces. For example, \citet{2007PhRvD..76d5002B} show the effect of varying the masses of the up and down quark, and find that star-and-chemistry permitting universes are huddled in a small shard of parameter space which has area $\Delta m_\textrm{up} \Delta m_\textrm{down} / m^2_\textrm{Planck} \approx 10^{-42}$.
\end{itemize}
Note that these constraints are all multi-dimensional; I have quoted one-dimensional bounds for simplicity. See \citet{2012PASA...29..529B} and references therein for plots demonstrating these and more constraints in multiple dimensions of parameter space. (It is has never been the case that the fine-tuning literature has varied one variable at a time.) 

These small numbers --- $10^{-110}$, $10^{-4}$, $10^{-33}$, $10^{-42}$ --- are, in the Bayesian fashion, an attempt to quantify our ignorance. We are not assuming the existence of a random universe-generating machine, nor describing the properties of a real or imagined statistical sample. The laws of nature as we know them contain arbitrary constants, which are not constrained by anything in theoretical physics. As usual, we can react to small probabilities in a couple of ways. Perhaps, like the probability of a deck of cards falling on the floor in a particular order, something improbable has happened. Enough said. Alternatively, like the probability that the burglar correctly guessed the 12-digit code by chance on the first attempt, it may indicate that we have made an incorrect assumption. We should look for an alternative assumption (or theory), on which the fact in question is not so improbable.

\subsection{Making Predictions in a Multiverse}

Theories are tested by their predictions, and we saw above that theory extensions are tested by their ability to predict the effective laws and constants of nature. In practice, this means calculating likelihoods.

The multiverse is an example of a "population plus selection effect" explanation. There is some observed outcome $X$ to be explained, and $X$ is highly improbable on any single trial. We postulate a large, varied population to explain why any $X$ exist at all, and a selection effect to explain why we observe $X$. For example, the front page of the newspaper reports correctly that Keith won the lottery. The probability of any particular person winning the lottery is very small. This occurrence is made more probable if we suppose that there are a large number of lottery players buying different tickets, and that only a lottery win would be considered newsworthy.

Where is the relevant selection effect when we are attempting to explain the statement that the effective laws of nature are $L$ and the associated free parameters are $\aU$, $\bU$? Recall that $U$ represents everything that \emph{I} know about \emph{this} universe. Thus, to explain $U$, the proposition $L \aU \bU$ must refer to this universe, the universe that I inhabit. $L \aU \bU$ cannot simply state that ``there is at least one universe in which the law $L$ holds and in which the constants are $\aU$, $\bU$'', because this will not explain the fact that I observe $U$.

This highlights an important difference in probability between calculating the probability that ``this X is Y'' and ``there is at least one X that is Y". Suppose I have just watched Alice deal herself five Royal flushes in a row in a game of poker. The probability of \emph{these} five hands being five Royal flushes assuming a fair deal is $10^{-29}$, making us wonder if Alice is cheating. The probability that \emph{someone}, somewhere has fairly dealt five Royal flushes depends on the number of poker deals there have ever been anywhere in the universe. If the universe is infinite, then this probability is one, making it useless for deciding whether Alice is cheating. As the Bayesian desiderata state, information must not be arbitrarily ignored. Reasoning as if we only knew that ``there is at least one instance of five Royal flushes'' is to discard information.

Note that the correct distinction is \emph{not} between first and third person probabilities, as is sometimes assumed in the multiverse literature. Third person probability can be as specific (``a particular X is Y'') as first person probabilities. Also, there is nothing ``mystical'' about using indexical information in probabilities \citep{2006math......8592N}; ``I'' can successfully select a particular individual -- in this case, the speaker of the sentence or calculator of the probability --- without assuming that the individual is unique in reality on account of ``some essence''.

So, what is the likelihood that \emph{this} universe has the observed constants, given a multiverse theory? We can calculate this in two pieces. We first calculate the probability is that observers exist  at all in the multiverse ($O$). So long as observer-permitting universes have non-zero chances and the universes in the multiverse are sufficiently varied, this probability will approach unity as the number of universes increases.
 
With an actual population of universes, the second probability piece is equal to a frequency: the fraction of observers (or observer moments) that observe our particular set of constants $\aU \bU$. This will depend on two factors: the rate $R_\textrm{obs}$ 
(per unit time and volume $\dd x^\mu$) at which observers/observations are made at particular point in spacetime, given the values of the ``constants'', and the probability of a particular set of constants at a particular spacetime point. Considering just the constants (\aL):
\begin{align} \label{eq:pmulti}
N_\textrm{obs} 		   &= \iint 		  			R_\textrm{obs}(x^\mu | \aL TLB) ~ p(\aL | x^\mu TLB) \dd x^\mu \df \aL \\
N_\textrm{obs} (\aU)  &= \iint\limits_{\aU} R_\textrm{obs}(x^\mu | \aL TLB) ~ p(\aL | x^\mu TLB) \dd x^\mu \df \aL \\
\Rightarrow p(\aU | OTLB) &= \frac{N_\textrm{obs} (\aU)} {N_\textrm{obs}}
\end{align}
The fine-tuning of the universe for intelligent life suggests that $R_\textrm{obs}$ is strongly peaked in our neighbourhood of parameter space, meaning that while regions of the universe with our constants are rare, they may be likely (or at least, not too unlikely) to be observed.

However, fine-tuning for life is not enough to ensure that a multiverse successfully predicts our constants of nature. The form of life with which we are familiar came about through biological evolution, via a gradual build up of complexity over timescales that are orders of magnitude longer than the lifetime of any particular individual. Such life forms require a stable planetary surface, a stable star producing usable photons, a ready supply of chemicals and so forth. However, observers could form without this history and environment as thermodynamic fluctuations. These \emph{Boltzmann Brains} can cause problem for a multiverse theory because they mean that $R_\textrm{obs}$ does not fall exactly to zero in seemingly hostile regions of parameter space.

In Equation \eqref{eq:pmulti}, we can write $R_\textrm{obs} = R_\textrm{life} + R_\textrm{BB}$ to represent the contribution of both biological life forms and Boltzmann Brains (BB) to the set of observers in a given multiverse. Thus, we can also write $N_\textrm{obs} = N_\textrm{life}  + N_\textrm{BB}$. We have, then, a competition between whether most observers (or observations) are made by common observers in rare conditions (life) or rare observers in common conditions (BB).

In testing a multiverse, it matters what other hypothetical observers in the multiverse observe, since the likelihood is normalized over $\aL$. Theories must place their bets as to what data are to be expected; for the multiverse, this means predicting what an observer will observe. While our calculation of the posterior involves evaluating the likelihood at our particular value of the constants in our universe, the normalization of the likelihood means that the more observers there are that do not observe what we observe, the smaller the likelihood. Every observer counts, not just those who observe exactly what we observe. 

\begin{figure}[t]
\centering
       \includegraphics[width=0.7\textwidth]{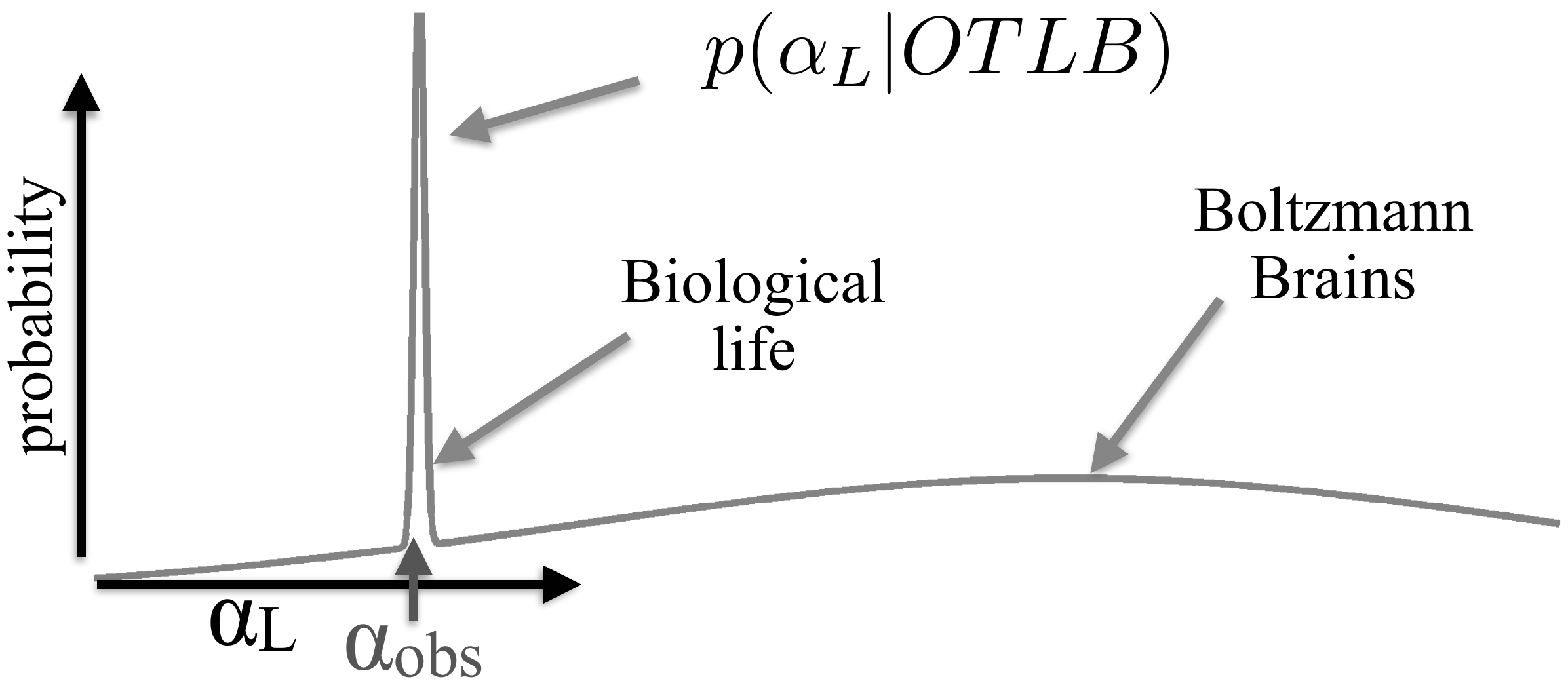}
	    \caption{An illustration of the Boltzmann observer problem. The likelihood of the set of constants that we observe given a multiverse theory $p(\alpha_L | OTLB)$ is normalised over $\alpha_L$. In evaluating the posterior probability of the multiverse, we evaluate the likelihood at the observed value of the constant $\alpha_\textrm{obs}$. Boltzmann brains can exist in universes which are hostile to biological life forms, and so can be found in a much larger region of parameter space. The larger the area under the broader Boltzmann Brain contribution, the smaller the (renormalized) likelihood of a biological life form observing $\alpha_\textrm{obs}$.} \label{fig:BBlikelihood}
\end{figure}

Figure \ref{fig:BBlikelihood} presents a 1D illustration of this \emph{Boltzmann observer problem}. The problem is \emph{not} that we might be Boltzmann Brains, or that most entities with my memories are fluctuation observers. We can call that the \emph{Boltzmann Me} problem, and set it to one side. The Boltzmann observer problem is a straightforward case of a failed prediction. A multiverse, once the full range of observers is considered, can be strongly disconfirmed by the seemingly innocuous observation that I am not a brain floating in empty space. The problem is not that we might be Boltzmann brains; the problem (for the theory) is that we aren't.

Testing the multiverse thus requires an understanding of the conditions under which observers can fluctuate into existence. It is of particular interest whether quantum fluctuations in a vacuum can create observers; see \citet{2014arXiv1405.0298B} for the case against such observers. In broadly thermodynamic terms, the Boltzmann Observer problem seems formidable. Biological life requires low entropy conditions in a large region; in fact, the entropy of this universe seems to be far lower than is required even by biological life forms \citep{Eddington1931,Penrose2004}. Boltzmann Brains, on the other hand, require only the smallest entropy fluctuation needed to create an observer. Given the usual connection between low entropy conditions and improbability, this would seem to make Boltzmann Brains far more numerous than biological life forms.

We also face the \emph{measure problem}, which in our formalism is the question of how to evaluate the likelihood of a multiverse theory when the number of observers is infinite. \citet[][p. 486-7]{Jaynes2003} warns that ``attempts to apply the rules of probability theory directly and indiscriminately on infinite sets'' leads to paradoxes, and that the only cure for this disease is that ``an infinite set should be thought of only as the limit of a specific (i.e. unambiguously specified) sequence of finite sets. \ldots The mathematically generated paradoxes have been found only when we tried to depart from this policy by treating an infinite limit as something already accomplished, without regard to any limiting operation.'' The problem for an infinite multiverse is that there is no such limit --- the infinity in question is ``completed'', an actually infinite set of universes and observers. In such circumstances, our probability assignments cannot be invariant under permutations of the labels on the observers \citep{2012PhRvD..86f3509O}. Infinite multiverse modellers could try to manufacture a limiting process --- perhaps a sequence of spacetime volumes --- or justify restricting attention to a finite subset. 

\subsection{Typicality and the Multiverse}

Testing the multiverse has often focused on \emph{typicality}: a theory is to be preferred if it predicts that human observers are typical in some class of objects in the universe \citep{2007PhRvD..75l3523H}. For example, suppose we derive from a multiverse theory $T$ the distribution of observed values of some constant $\alpha$: $p(\alpha|TB)$. $T$ predicts that, with 95\% certainty, our observed value of $\alpha$ falls inside the central 95\% of the distribution. If this prediction is correct, then the theory has passed this test.

This type of reasoning is transparently frequentist: the only probabilities that we can define are those of data with respect to theory, so we test theories by inventing a test for the likelihood. Should it pass, we try to think of another test, or else get more data. It ignores prior probabilities, and so cannot calculate the probability \emph{of} a theory given the evidence.

As with other frequentists methods, we can use Bayesian probability theory to expose the hidden assumptions. When is typicality --- defined as closeness to the likelihood peak --- a useful discriminant between models? Consider the simple case of two theories $T_1$ and $T_2$ competing to predict the value of some constant $\alpha$. We calculate the likelihood distribution for $\alpha$ on each theory $p(\alpha|TB)$; suppose that it is roughly Gaussian. If a) the prior probabilities of $T_1$ and $T_2$ are similar and, b) if the widths of the likelihood distributions $p(\alpha|T_1 B)$ and $p(\alpha|T_2 B)$ are similar, then the theory for which the observed value of $\alpha$ lies closest to the peak of the distribution has the greater posterior probability.

Note that both conditions a) and b) are needed, and thus typicality is neither a necessary nor a sufficient condition for a multiverse theory to be a good theory. The problem with typicality is that it compares values of the likelihood at different values of $\alpha$, when we should be comparing different theories by evaluating their likelihoods at the observed value of $\alpha$.

Let's be clear of the status of typicality. It is not an assumption to be accepted or rejected at our leisure. It is not an assumption at all. Under certain conditions, it is useful rule of thumb in evaluating competing multiverse theories. Bayesianism identifies these conditions.

\section{Extension 3: Getting Metaphysical}

At this conference, George Ellis has invited us to think about not only cosmology with a small `c', defined as the the physics of the universe on large scales, but also Cosmology with a capital `C', which asks the great questions of existence, meaning and purpose that are raised by physical cosmology. Nothing in our formalism assumes that $T$ is a \emph{physical} theory. Indeed, if there is a final, ultimate physical theory of nature $F$, then whatever we think \emph{about} that theory will have to be deeper than physics, so to speak. Even if all that remains is to state the definition of \emph{naturalism}, that nothing other than the physical exists, we must acknowledge that this is a statement about physics, not of physics.

Further,we want to know whether or not naturalism is true. We can treat naturalism like any other theory, and consider its prior probability $p(N|B)$, and the probability of the final scientific laws on naturalism $p(F|NB)$. Even if we can't calculate these quantities, they point to the right questions to ask. Naturalism, as a hypothesis, is what statisticians call \emph{non-informative} --- it gives us no reason to prefer any particular $F$. In the case of naturalism, this is an \emph{in principle} ignorance, since by hypothesis there are no true facts that explain why $F$ rather than some other final law, why any law at all, why a mathematical law, what ``breathes fire into the equations and makes a universe for them to describe?'' \citep{1988bhtb.book.....H}, what is existence, and so on.

Non-informative theories have likelihoods that are at the mercy of the size of their possibility space. For example, ``the burglar guessed the 12-digit security code'' gives us no reason to prefer any code over any other, and thus the likelihood of any particular code should reflect these trillion possibilities. The only thing in our background knowledge $B$ that restricts the set of possible universes is internal (mathematical) consistency. Naturalism, then, is at the mercy of every possible way that concrete reality could consistently be. This places naturalism in an unenviable position.

Its competitors to explain $F$ include axiarchism \citep{1989univ.book.....L} and theism \citep{Swinburne2004}, which argue that we should expect the existence of physical reality with significant \emph{moral} value, including the moral good of embodied, free, conscious moral agents. Axiarchism and theism, then, bet heavily on the subset of possible laws that permit the existence of such life forms. Whether the fine-tuning of the laws \emph{as we know them} ($\Lab$) for life extends to final laws $F$, and their relative prior probabilities, will decide whether any of these theories is preferable to naturalism.

\section*{Acknowledgments}
I would like to thank all the attendees of the Philosophy of Cosmology UK/US Conference, 2014, Tenerife for stimulating talks and discussions. Supported by a grant from the John Templeton Foundation.  This publication was made possible through the support of a grant from the John Templeton Foundation. The opinions expressed in this publication are those of the author and do not necessarily reflect the views of the John Templeton Foundation.


\begin{thebibliography}{99}

\bibitem[Adams(2008)]{2008JCAP...08..010A} Adams, F.~C.\ (2008), Journal of Cosmology and Astroparticle Physics, 8, 010 

\bibitem[Barnes(2012)]{2012PASA...29..529B} Barnes, L.~A.\ (2012), Publications of the Astronomical Society of Australia, 29, 529

\bibitem[Barr \& Khan(2007)]{2007PhRvD..76d5002B} Barr, S.~M., \& Khan, A.\ (2007), Physical Review D, 76, 045002 

\bibitem[Boddy et al.(2014)]{2014arXiv1405.0298B} Boddy, K.~K., Carroll, 
S.~M., \& Pollack, J.\ (2014), E-print arXiv:1405.0298 

\bibitem[Cameron 
\& Pettitt(2012)]{2012arXiv1207.6223C} Cameron, E., \& Pettitt, T.\ (2012), E-print arXiv:1207.6223 

\bibitem[\protect\citeauthoryear{Caticha}{2009}]{Caticha2009} Caticha, A. (2009) ``Quantifying Rational Belief.'' AIP Conf. Proc. 1193, 60

\bibitem[\protect\citeauthoryear{Eddington}{1931}]{Eddington1931} Eddington, A. S. (1931) ``The End of the World: from the Standpoint of Mathematical Physics'' Nature 127, 3203

\bibitem[\protect\citeauthoryear{Einstein}{1949}]{Einstein1949} Einstein, A. (1949) ``Autobiographical Notes'', in Schilpp, P. A. (Ed.) Albert Einstein, Philosopher-Scientist. Open Court Publishing Company, Illinois

\bibitem[\protect\citeauthoryear{Fisher}{1921}]{Fisher1921} Fisher, R. A. (1921) ``On the `Probable Error' of a Coefficient of Correlation Deduced from a Small Sample.'' Metron, 1: 3–32. 162, 164

\bibitem[\protect\citeauthoryear{Glymour}{1980}]{Glymour1980} Glymour, C. (1980) ``Theory and Evidence''. Princeton: Princeton University Press

\bibitem[\protect\citeauthoryear{Griffiths}{2008}]{Griffiths2008} Griffiths, D. (2008) ``Introduction to Elementary Particles''. New York: John Wiley \& Sons

\bibitem[Hartle \& Srednicki(2007)]{2007PhRvD..75l3523H} Hartle, J.~B., \& Srednicki, M.\ (2007), Physical Review D, 75, 123523 

\bibitem[Hawking(1988)]{1988bhtb.book.....H} Hawking, S.~W.\ (1988), Toronto: 
Bantam Books

\bibitem[\protect\citeauthoryear{Jaynes}{2003}]{Jaynes2003} Jaynes, E. T. (2003) ``Probability Theory: The Logic of Science.'' Cambridge University Press, Cambridge, UK

\bibitem[King et al.(2012)]{2012MNRAS.422.3370K} King, J.~A., Webb, J.~K., 
Murphy, M.~T., et al.\ (2012), Monthly Notices of the Royal Astronomical Society, 422, 3370 

\bibitem[\protect\citeauthoryear{Kempthorne \& Folks}{1971}]{Kempthorne1971} Kempthorne, Oscar, \& Folks, Leroy. ``Probability, statistics, and data analysis''. Ames, Iowa: The Iowa State University Press

\bibitem[\protect\citeauthoryear{Kolmogorov}{1933}]{Kolmogorov1933} Kolmogorov, A. N. (1933). Translated as ``Foundations of Probability'', New York: Chelsea Publishing Company (1950)

\bibitem[Leslie(1989)]{1989univ.book.....L} Leslie, J.\ (1989), London, New 
York: Routledge

\bibitem[\protect\citeauthoryear{MacKay}{2003}]{MacKay2003} MacKay, D. J. C. (2003) ``Information Theory, Inference, and Learning Algorithms''. Cambridge: Cambridge University Press

\bibitem[Neal(2006)]{2006math......8592N} Neal, R.~M.\ (2006), 
E-print arXiv:math/0608592 

\bibitem[Olum(2012)]{2012PhRvD..86f3509O} Olum, K.~D.\ (2012), Physical Review D, 86, 
063509 

\bibitem[\protect\citeauthoryear{Penrose}{2004}]{Penrose2004} Penrose, R. (2004) ``The Road to Reality: A Complete Guide to the Laws of the Universe''. London: Jonathan Cape

\bibitem[\protect\citeauthoryear{Planck Collaboration et al.}{2015}]{Planck2015} Planck Collaboration et al. Ade, P.~A.~R., Aghanim, N., et al.\ 2015, arXiv:1502.01589

\bibitem[\protect\citeauthoryear{Swinburne}{2004}]{Swinburne2004} Swinburne, R. (2004) ``The Existence of God''. Oxford: Oxford University Press

\bibitem[Tegmark 
\& Rees(1998)]{1998ApJ...499..526T} Tegmark, M., \& Rees, M.~J.\ (1998), The Astrophysical Journal, 499, 526 

\bibitem[Webb et al.(2011)]{2011PhRvL.107s1101W} Webb, J.~K., King, J.~A., 
Murphy, M.~T., et al.\ (2011), Physical Review Letters, 107, 191101 

\bibitem[\protect\citeauthoryear{Weinberg}{1993}]{Weinberg1993} Weinberg, S. (1993). ``Dreams of a final theory''. London: Vintage

\bibitem[Whitmore 
\& Murphy(2015)]{2015MNRAS.447..446W} Whitmore, J.~B., \& Murphy, M.~T.\ (2015), Monthly Notices of the Royal Astronomical Society, 447, 446 

\end{thebibliography}
\end{document}